\documentclass[aps,prb,floatfix,twocolumn]{revtex4}%
\usepackage{amsmath}
\usepackage{graphicx}
\usepackage{amsfonts}
\usepackage{amssymb}%
\providecommand{\U}[1]{\protect\rule{.1in}{.1in}}
\providecommand{\U}[1]{\protect\rule{.1in}{.1in}}
\providecommand{\U}[1]{\protect\rule{.1in}{.1in}}
\providecommand{\U}[1]{\protect\rule{.1in}{.1in}}

\begin{document}
\title{Scaling laws of the Kondo problem at finite frequency}
\author{L.E Bruhat$^{1,2}$\footnote{%
To whom correspondence should be addressed: bruhat@chalmers.se}, J.J.
Viennot$^{1,3}$, M.C. Dartiailh$^{1}$, M.M. Desjardins$^{1}$, A. Cottet$^{1}$, T. Kontos$^{1}$\footnote{%
To whom correspondence should be addressed: kontos@lpa.ens.fr}}
\affiliation{$^{1}$Laboratoire Pierre Aigrain, Ecole Normale Sup\'{e}rieure-PSL Research University, CNRS, Universit\'{e}s Pierre et Marie Curie-Sorbonne Universit\'{e}s, Universit\'{e} Denis Diderot-Sorbonne Paris Cit\'{e}, 24, rue Lhomond, 75231 Paris Cedex 05, France\\}
\affiliation{$^{2}$Microtechnology and Nanoscience, Chalmers University of Technology, Kemivgen 9, SE-41296 Gothenburg, Sweden\\}
\affiliation{$^{3}$JILA and Department of Physics, University of Colorado, Boulder, Colorado, 80309, USA\\
}
\date{\today}

\begin{abstract}
Driving a quantum system at finite frequency allows one to explore
its dynamics. This has become a well mastered resource for controlling the
quantum state of two level systems in the context of quantum information
processing. However, this can also be of fundamental interest, especially
with many-body systems which display an intricate finite frequency behavior.
In condensed matter, the Kondo effect epitomizes strong electronic
correlations, but the study of its dynamics and the related scaling laws has
remained elusive so far. Here, we fill this gap by studying a carbon
nanotube based Kondo quantum dot driven by a microwave signal. Our
findings not only confirm long-standing theoretical predictions, but also
allow us to establish a simple ansatz for the scaling laws on the Kondo
problem at finite frequency. More generally, our technique opens a new path
for understanding the dynamics of complex quantum dot circuits in the
context of quantum simulation of strongly correlated electron fluids.
\end{abstract}
\maketitle

\section{INTRODUCTION}

The use of quantum dot circuits to study the physics of interacting electrons in a
controlled way is now well established. Such
nanoscale circuits are ideally suited to study correlation effects, thanks to their tunability \cite
{Goldhaber-Gordon1998a,Nygard2000,Wiel2000,Iftikhar2015,Keller2015} and
natural interface with high frequency signals \cite{Desjardins2017}. Correlation effects are
deeply related to Kondo physics which remains a major topic of condensed matter.
While there has been a lot of experiments related to this phenomenon in the static regime, there are only few experiments
in the dynamic regime, which can be accessed for example through the study of electronic transport under high frequency driving.

The study of high frequency driven quantum dots is almost as old as the
advent of quantum dot circuits, due to the interest on photon assisted
tunneling (PAT). Most of the works have focused on the Coulomb blockade
regime \cite{Kouwenhoven1994a,Wiel2003}, for which Coulomb conductance peaks
appear when dot charge fluctuations are energetically allowed. In this case,
PAT processes essentially produce side-bands around the Coulomb peaks,
located at source-drain bias voltages $V_{SD}\approx \pm hf /e$\ where
$f $\ is the frequency of the microwave radiation and $e$\ is the
elementary charge. These experiments, which are well understood, have even
been generalized to THz frequencies or to on-chip detection schemes for
quantum noise \cite{Shibata2012,Deblock2003,Onac2006}.

In contrast, on the experimental side, PAT has raised little attention in strongly correlated quantum dot circuits. However, strong interactions lead to a rich dynamical
phenomenology\cite%
{Ng1996,Konig1996,Nordlander1998,Lopez1998,Goldin1998,Schiller1996,Kaminski1999,Kaminski2000,Beri2012}%
. A radiation at frequency $f$\ is a natural tool to study such finite
frequency processes. For instance the Kondo dynamics, which depends on the
Kondo temperature $T_{K}$, could be scrutinized by using $k_{B}T_{K}\sim hf$%
, in principle. However, even the simple question of whether or not
side-bands appear around the Kondo resonance peak\ is intricate. Whereas
there is a rich theoretical literature \cite%
{Ng1996,Nordlander1998,Lopez1998,Schiller1996,Kaminski1999,Kaminski2000},
only four experiments have been carried out on that topic in the last decade%
\cite{Elzerman,Kogan2004b,Yoshida2015,Hemingway2014a}. Noticeably,
contradicting results were found on the existence\cite%
{Kogan2004b,Yoshida2015} or not\cite{Elzerman} of microwave induced
side-bands for a Kondo peak. In addition, the predictions of a scaling
behaviour\cite{Kaminski2000}, which is the hallmark of the Kondo effect\cite%
{Anderson1976,Nygard2000,Wiel2000} could not be tested experimentally\cite{Hemingway2014a}. Strikingly,
although the dot microscopic description depends on many parameters, its
reduced zero voltage conductance is expected to depend only on the reduced
amplitude $eV_{AC}/k_{B}T_{K}$ and frequency $hf/k_{B}T_{K}$\ of the
oscillating voltage applied to the dot.

This lack of quantitative test is due to the experimental
challenge of controlling both the excitation amplitude $V_{AC}$ and the
excitation frequency $f$ over a broad band. To cope with these difficulties,
previous experimental studies of the AC Kondo effect either worked at a
fixed frequency\cite{Kogan2004b}, or partly postulated the Kondo response
itself to justify their microwave amplitude calibration\cite{Elzerman,
Hemingway2014a}. These facts probably explain the lack of clear consensus on
the underlying physics of photon assisted tunneling of a Kondo dot.

The key achievement of our work is precisely to provide the first
quantitative amplitude-frequency characterization of the AC Kondo effect. We
use a carbon nanotube based quantum dot exposed to a microwave radiation. We
find that the existence of photon-assisted tunneling peaks for the AC Kondo
problem depends on the excitation frequency used. In contrast to previous
works, we use an independent calibration method for $V_{AC}$, based on the
adiabatic response of the current in the Coulomb blockade regime (see Appendix B).
This allows us to study the full amplitude-frequency map of the Kondo peak at zero bias, and to make a
quantitative test of the long-standing theoretical predictions of Kaminski,
Nazarov and Glazman \cite{Kaminski2000}. In particular, we observe an
apparently paradoxical prediction of the theory: in the case of bias
excitation, the higher the frequency $f$, the lower the effect on the Kondo
peak at a constant microwave amplitude $V_{AC}$. We finally compare two
Kondo peaks with different $T_{K}$'s, in order to perform a full test of the
scaling properties of the Kondo effect at finite frequency. We are able to
test simple analytical expressions for the scaling properties of the Kondo
resonance using an ansatz adapted from the work of Kaminski, Nazarov and
Glazman.

{\small
\begin{figure}[tph]
{\small \centering\includegraphics[width=1\linewidth,angle=0]{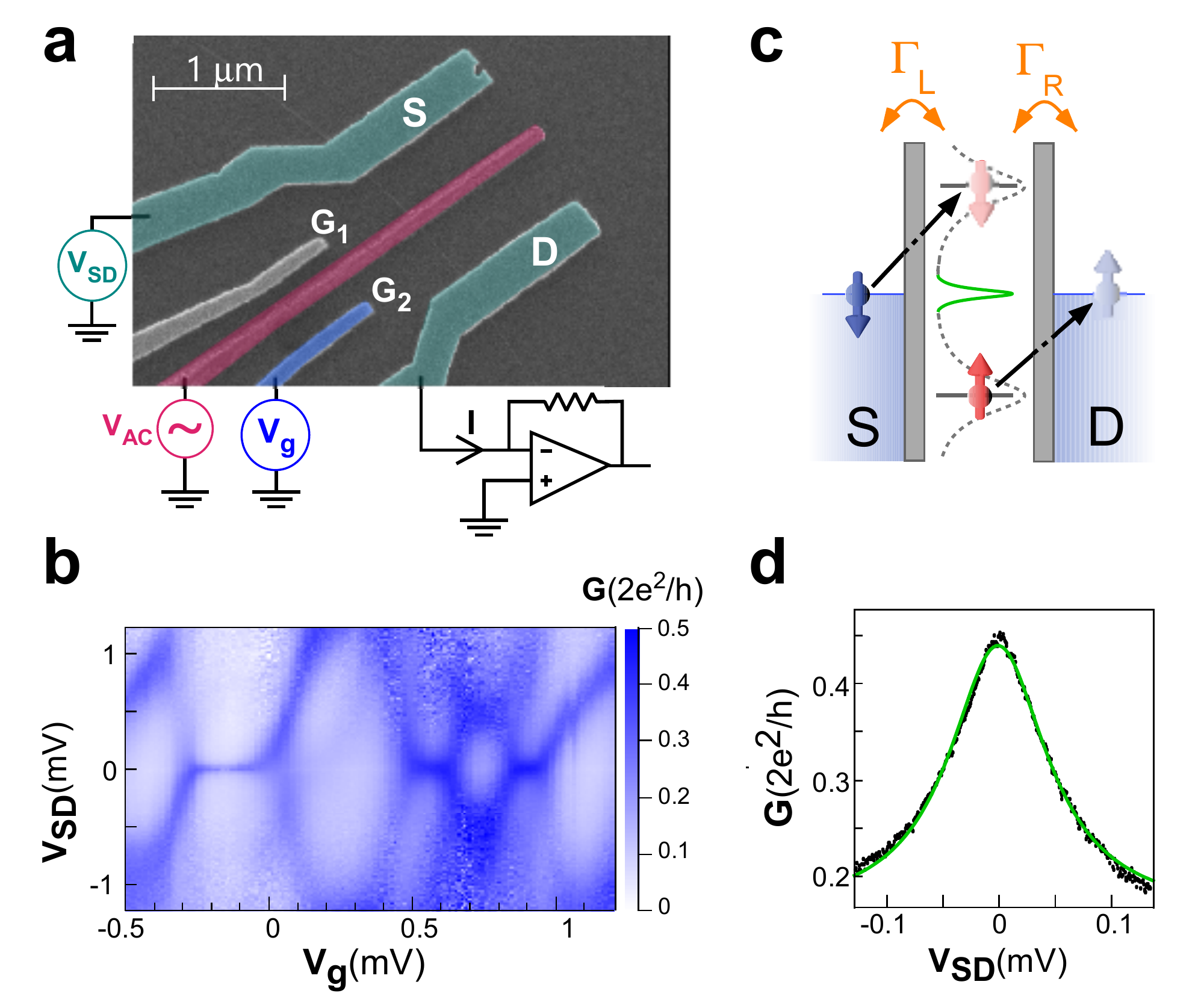}  }
\caption{a. SEM picture of our device. It behaves as a single dot, which can be
biased by applying a dc voltage $V_{SD}$ on the source electrode, while the
current I is measured through the drain electrode. A DC gate voltage $V_{G}$
is applied on a side gate. A microwave excitation $V_{AC}$ at frequency $f$
can be applied on the central top gate. b. Conductance colour plot versus
gate and bias voltage showing characteristic Kondo ridges. c Example of
second order tunnel process which flips the dot spin. The constructive
intereference of such processes at all orders give rise to the Kondo peak
for $V_{SD}=0$. d. Plot of Kondo resonance conductance peak at
equilibrium ($V_{AC}=0$). The black dots are data and the green solid line
is a Lorentzian fit giving $k_{B}T_{K}=32.7\protect\mu eV$, $%
G_{OFF}=0.281(2e^{2}/h)$, $G_{BG}=0.159(2e^{2}/h)$ (see equation \protect\ref%
{equation:KondoAC:G_eq_fit} in Appendix A).}
\label{fig:sample}
\end{figure}
}

\section{EXPERIMENTAL SETUP}

Our device is a carbon nanotube quantum dot circuit embedded in a coplanar
waveguide microwave cavity (figure \ref{fig:sample}a). A bias voltage $%
V_{SD} $ can be applied to one of the two Pd contact electrodes (emerald
green). A DC gate voltage $V_{g}$ applied on a side gate electrode (blue)
tunes the dot orbital energy. A central top gate electrode (pink) is used to
apply a microwave signal to the device. Our calibration shows that although it is connected to the top gate electrode,
it predominantly results in a source and drain AC bias voltage $V_{AC}$ (see Appendix). Our setup measures the
current I and the DC differential conductance $G=dI/dV_{SD}$ at 20mK. Figure %
\ref{fig:sample}b shows a $V_{SD}-V_{g}$ conductance color plot in absence
of microwave excitation ($V_{AC}=0$). It displays zero-bias ridges
characteristic of the Kondo regime. One of the Kondo resonance conductance
peak (figure \ref{fig:sample}d) corresponds to a Kondo temperature $%
T_{K}\sim 380mK$ extracted from a Lorentzian fit (see Appendix A). Note that
typical Kondo temperatures in carbon nanotubes rather lie in the $1-10K$
range, but a lower $T_{K}$ is advantageous to explore the Kondo dynamics in
the relevant regime $k_{B}T_{K}\sim hf$ with accessible GHz frequencies.

\section{KONDO PEAK UNDER MICROWAVE IRRADIATION}

{\small
\begin{figure}[!htbp]
{\small \centering\includegraphics[width=1.\linewidth,angle=0]{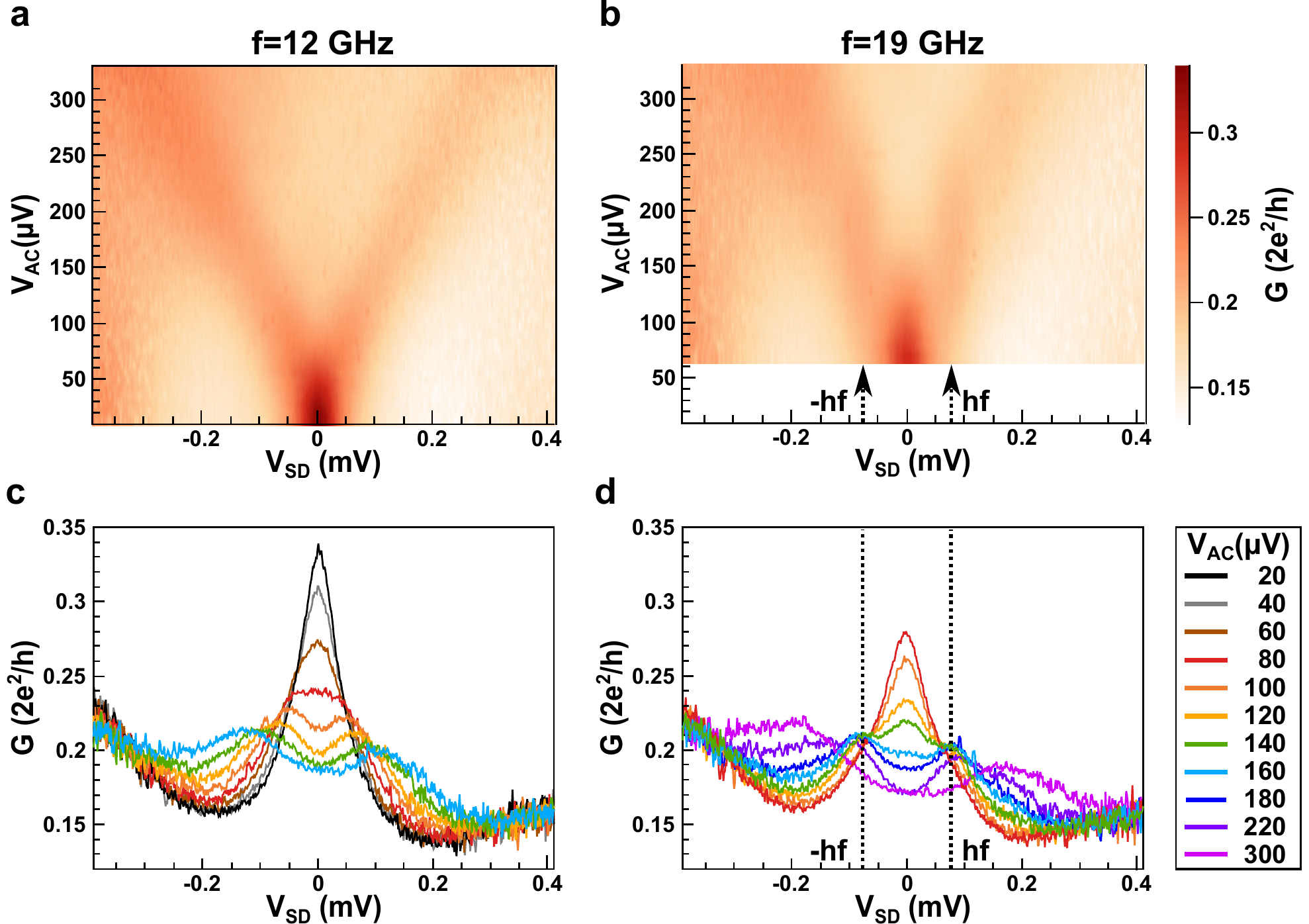}
}
\caption{a. (resp. b.) Colour plot of the conductance versus bias voltage $V_{SD}$
and excitation voltage $V_{AC}$ with excitation frequency $f=12~$\textrm{GHz}
(resp. $f=19~$\textrm{GHz}) corresponding to reduced frequencies $hf/k_{B}T_{K}=1.8$
(resp. $hf/k_{B}T_{K}=2.8$). $k_{B}T_{K}$, $G_{OFF}$ and $G_{BG}$ for this data set differ slightly from the data shown in Figure \ref{fig:sample}b and their values are given in the Table (figure 8, file number 18). c. (resp. d.) Linear cuts of data shown in upper panel a.
(resp. b.). The lines colour encodes the excitation voltage, and the colour
code applies to both frequencies.}
\label{fig:ACphenomenology}
\end{figure}
}

We now apply a finite microwave excitation $V_{AC}$ and study how the Kondo
resonance conductance peak is modified, for two
excitation frequencies $f=12GHz$ and $f=19GHz$ for which the microwave
excitation is dominantly coupled to the source-drain bias of the dot (see Appendix A). Our
measurements confirm the phenomenology observed in previous experiments
(figure \ref{fig:ACphenomenology}). The strongest the microwave excitation,
the lowest the Kondo resonance peak (figure \ref{fig:ACphenomenology}c,d).
Above a certain excitation voltage, several peaks are visible. For $f=12GHz$%
, we observe a linear splitting of the Kondo resonance versus $V_{AC}$
(figure \ref{fig:ACphenomenology}a,c), which corresponds to the adiabatic
regime. In contrast, for $f=19GHz$, three peaks can be resolved
simultaneously (green curve on figure \ref{fig:ACphenomenology}d), and the
two side peaks remain locked at $\pm hf$ over a certain range of excitation
voltage (figure \ref{fig:ACphenomenology}b,d). These satellites are the
manifestation of photo-induced Kondo resonances. Figures \ref%
{fig:ACphenomenology}c and \ref{fig:ACphenomenology}d also show that the
conductances at the Kondo resonance ($V_{SD}=0$) with and without applied microwave excitation ($G_{ON}$ and $G_{OFF}$) clearly differ and
that the variation of $G_{ON}$ with $V_{AC}$ quantitatively differs for $%
f=12GHz$ and $f=19GHz$.

{\small
\begin{figure}[tbp]
{\small \centering\includegraphics[width=1.\linewidth,angle=0]{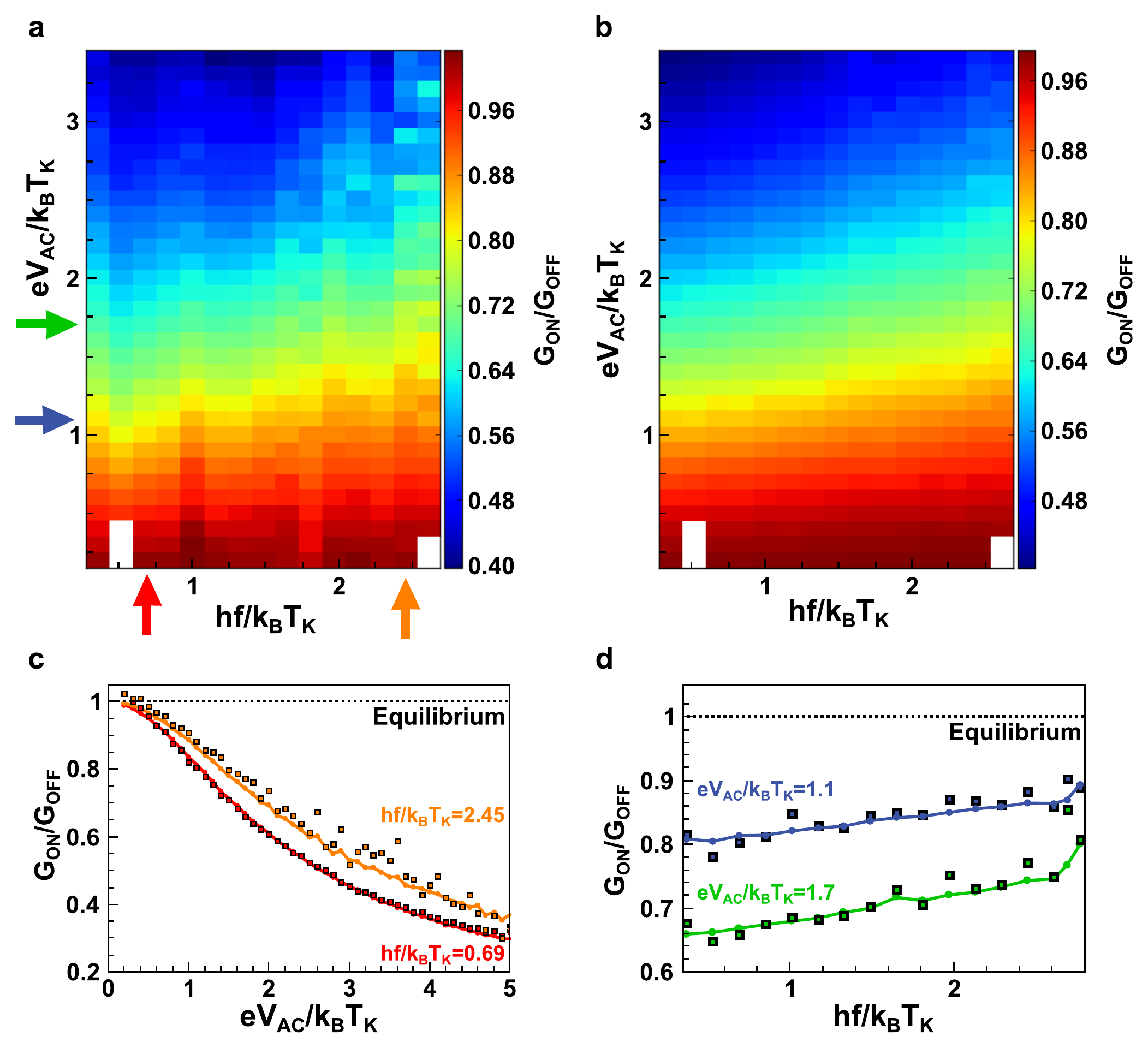}
}
\caption{a. Experimental averaged normalized conductance colour plot $G_{ON}/G_{OFF}$
as a function of excitation scaled amplitude $eV_{AC}/k_{B}T_{K}$ and scaled
frequency $hf/k_{B}T_{K}$. b. Calculated map of $G_{ON}/G_{OFF}$ using the
Ansatz formula \protect\ref{equation:KondoAC:Ansatz}. c. Linear cuts at $%
hf/k_{B}T_{K}$=0.69 (red) and $hf/k_{B}T_{K}$=2.45 (orange). d. Linear cuts
at $eV_{AC}/k_{B}T_{K}$=1.1 (blue) and $eV_{AC}/k_{B}T_{K}$=1.7 (green). In
panels c. and d. squares correspond to data points, while dots linked by
solid lines correspond to the Ansatz. The black dashed line indicates the
equilibrium reference ($V_{AC}=0$).}
\label{fig:FVmap}
\end{figure}
}

Using our calibration, we obtain the experimental map of the conductance
ratio $G_{ON}/G_{OFF}$ with and without microwave excitation as a function
of the scaled amplitude $eV_{AC}/k_{B}T_{K}$ and scaled frequency $%
hf/k_{B}T_{K}$ (figure \ref{fig:FVmap}a, see Appendix A). While $%
G_{ON}/G_{OFF}$ decreases with $V_{AC}$, it increases with $f$. These trends
appear even more clearly on the map line cuts plotted in figure \ref%
{fig:FVmap}c and \ref{fig:FVmap}d. The higher the amplitude, the further the
conductance at the Kondo resonance gets away from the equilibrium value, up
to data noise. In contrast, the higher the frequency, the closer the Kondo
conductance gets to its equilibrium value. Both behaviors were predicted by
Kaminski, Nazarov and Glazman for an AC excitation of the bias voltage, in
the limits $f\rightarrow 0$ and $V_{AC}\rightarrow 0$ respectively. From
their theoretical work, we build the following Ansatz to quantitatively
describe our data for finite values of $f$ and $V_{AC}$:
\begin{widetext}
\begin{equation}
\dfrac{G_{ON}}{G_{OFF}}\left( \dfrac{eV_{AC}}{k_{B}T_{K}},\dfrac{hf}{k_{B}T_{K}},G_{OFF}\right) =\dfrac{1}{\sqrt{1+\dfrac{3}{8}%
\left( \dfrac{eV_{AC}}{k_{B}T_{K}}\right) ^{2}\alpha \left( \dfrac{hf}{k_{B}T_{K}},G_{OFF}\right)}}  \label{equation:KondoAC:Ansatz}
\end{equation}
\end{widetext}
with
\begin{equation}
\alpha \left( \dfrac{hf}{k_{B}T_{K}},G_{OFF}\right) =\dfrac{1}{\sqrt{1+\left( \dfrac{16}{3}\dfrac{a}{\pi }%
\dfrac{G_{OFF}}{2e^{2}/h}\dfrac{k_{B}T_{K}}{hf}\right) ^{-2}}}
\end{equation}%
where $a\sim 1$ is a dimensionless parameter. More precisely, this
expression bridges the analytical expressions of $G_{ON}/G_{OFF}$ in the
adiabatic limit (Eq.(76) of Ref.\cite{Kaminski2000}) and in the
low-amplitude high-frequency limit (Eq.(74) of Ref.\cite{Kaminski2000}).
Strikingly, our Ansatz is able to reproduce the full amplitude-frequency map
of $G_{ON}/G_{OFF}$ with a single adjustable parameter a=4.4 (figure \ref%
{fig:FVmap}b). A 10$\%$ global correction has been applied to amplitudes ($%
\tilde{V}_{AC}=1.1V_{AC}$), to account for absolute calibration uncertainty.
The good agreement between our Ansatz and our measurements can be better
appreciated from the linear cuts displayed in \ref{fig:FVmap}c and \ref%
{fig:FVmap}d.

Can we find a physical picture for the counter-intuitive frequency
dependence we observe? Indeed one would rather think that higher frequencies
would represent more energetic perturbations for the Kondo cloud, thus
resulting in more decoherence. However, our AC excitation affects the phase
difference $\varphi $ between the electronic states in the source and drain
electrodes, which becomes\cite{Tien1963} $\varphi (t)=e/\hbar \int^{t}{%
V_{AC}\cos {(2\pi f{t}^{\prime })}\mathrm{d}{t}^{\prime }=eV_{AC}\sin {(2\pi
f{t})}/hf}$ for ${V_{DC}=0}$. These phase oscillations trigger inelastic
cotunneling processes in which the dot spin is flipped and an electron is
sent above the Fermi level of the reservoirs, so that the circuit gets out
of the highly conducting many-body singlet Kondo ground state (see Figs.\ref%
{fig:QualExp}a and d). The rate $\Gamma _{e}$ associated to this decoherence
process depends on the amplitude of the phase oscillations, which vanish at
high excitation frequencies due to the $1{/f}$ factor in the above equation
(see Fig.\ref{fig:QualExp}b). This simple fact explains that the Kondo cloud
is less affected by higher frequencies. More quantitatively, at first order
in ${eV_{AC}/hf}$, the two spin states of the dot are coupled by an oscillating
matrix element with amplitude $J_{0}e{V_{AC}/hf}$ (see Eq.(18) of Ref.\cite{Kaminski2000}),
with $J_{0}$ the spin
exchange factor of the Kondo hamiltonian in equilibrium. In the limit $%
\hbar \Gamma _{e}\ll k_{B} T_{K}$, which we consider for simplicity, decoherence is
dominated by single photon absorption processes such as the one represented
in \ref{fig:QualExp}a, and it involves initial reservoir states which must
lie in a band of width ${f}$ below the Fermi energy of the leads, as visible
from Fig.\ref{fig:QualExp}a. Hence, from the Fermi's golden rule, one gets $%
\Gamma _{e}=(J_{0}\nu e{V_{AC})}^{2}{/\hbar }^{2}{f}$ with $\nu $ the
density of states in the leads. Once the dot spin is flipped, its
conductance is negligible in comparison with the unitary conductance $G_{U}$
of the Kondo limit. Since the circuit relaxes to the Kondo ground state with
a rate $k_{B}T_{K}/\hbar$, the average conductance reads $G=G_{U}/(1+\hbar%
\Gamma _{e}/k_{B}T_{K})\simeq (1-\alpha {V_{AC}^{2}/f}T_{K})G_{U}$, with $\alpha
=2\pi J_{0}^{2}e^{2}\nu ^{2}{/h k}_{B}$, which corresponds to Eq.(74) of Ref.\cite{Kaminski2000}.
This reasoning explains the increase of $G$
with $f$ but also agrees with the decrease of $G$ with ${V_{AC}}$.

{\small
\begin{figure}[tbp]
{\small \centering\includegraphics[width=1.\linewidth,angle=0]{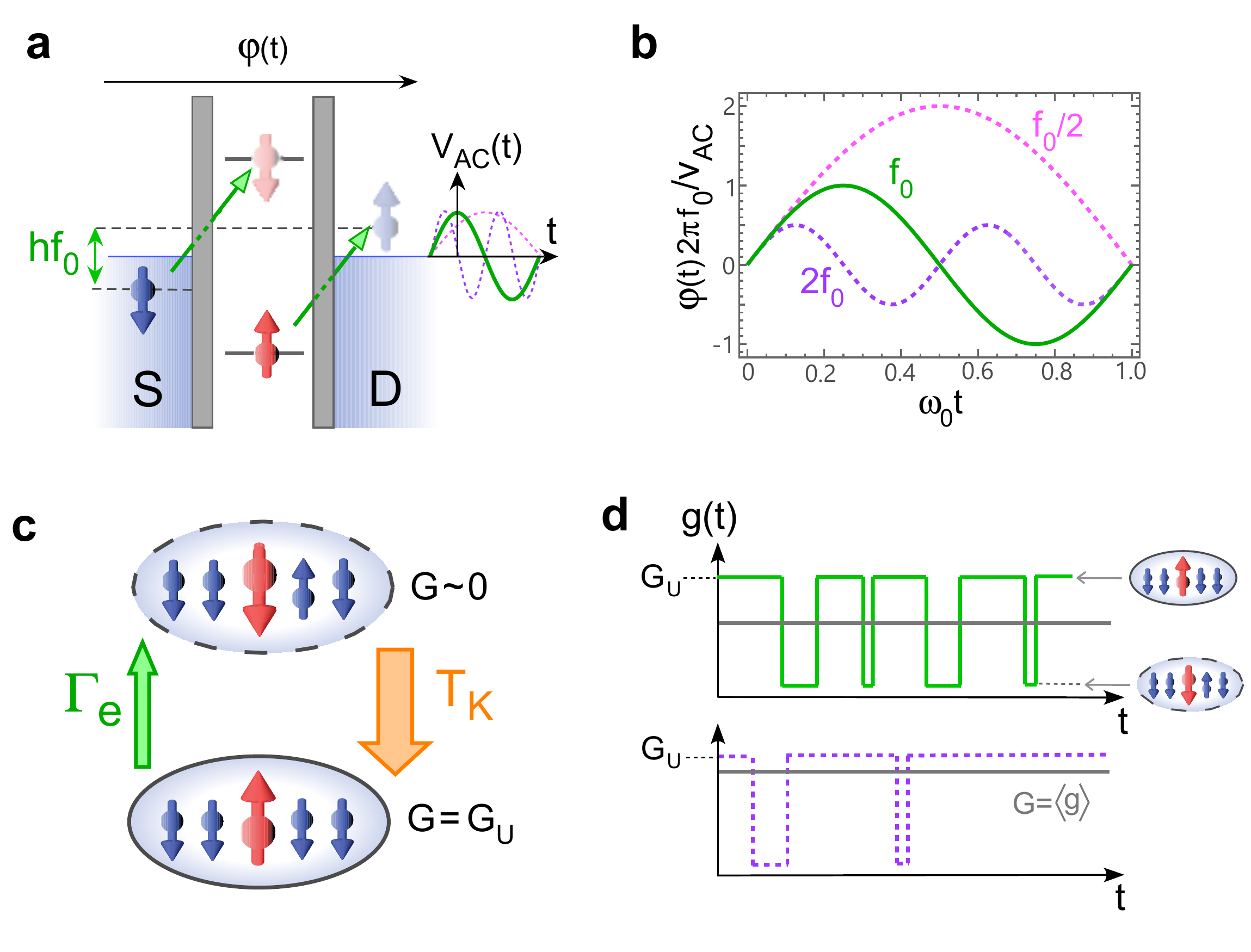}
}
\caption{a. Sketch of a quantum dot with an AC bias voltage $V_{AC}(t)=V_{AC}\cos (%
\protect2\pi ft)$. Due to this modulation, photons can trigger second order
tunnel processes, represented here by the pair of green arrows, which flip
the dot spin and leave a high energy electron in the reservoirs. The three
colored solid/dotted lines represent $V_{AC}(t)$ for different values of
excitation frequency f with the same $V_{AC}$. This color
code applies to all panels. b. Plot of the phase difference $\protect\varphi %
(t)$ between the electronic states in the source and drain, as a function of
time. c. Representation of the circuit dynamics. The dot leaves the Kondo
ground state with a rate $\Gamma _{e}$ and relaxes to the Kondo ground state
again with a rate $k_B T_{K}/\hbar$. d. Time dependence of the dot conductance $g(t)$%
, which reaches $G_{U}$ in the Kondo state and is much smaller otherwise.
The average conductance $G=\left\langle g(t)\right\rangle $ measured in the
experiment is closer to $G_{U}$ for large values of f for
which $\Gamma _{e}$ is smaller.}
\label{fig:QualExp}
\end{figure}
}

\section{SCALING LAWS}

{\small
\begin{figure}[tbp]
{\small \centering\includegraphics[width=1.\linewidth,angle=0]{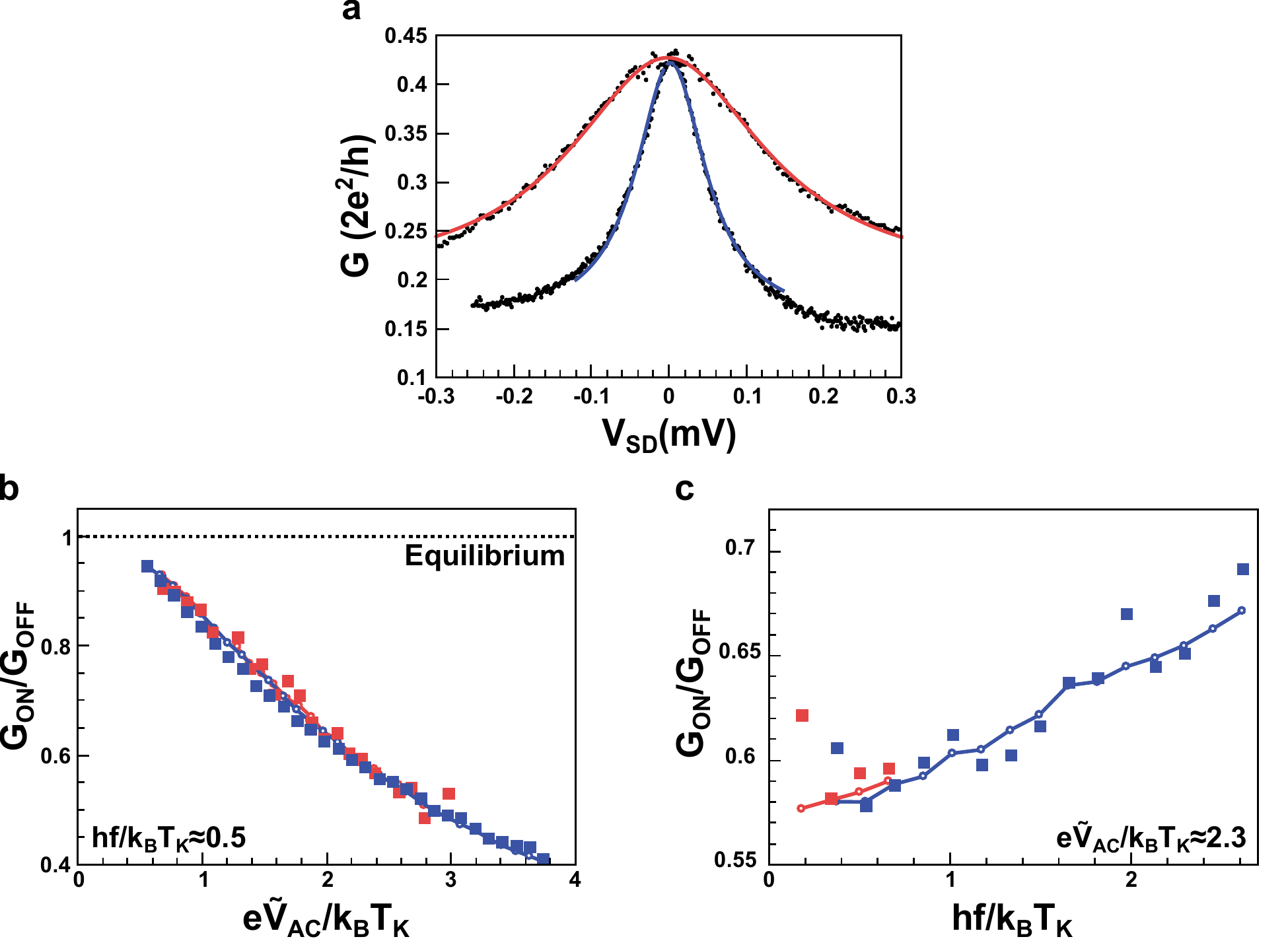}
}
\caption{a. Plots of conductance versus bias voltage for narrow and broad Kondo
resonances at equilibrium ($V_{AC}=0$). The black dots are data and the
solid lines are Lorentzian fits yielding $k_{B}T_{K}=33.3\protect\mu eV$ for
the narrow Kondo peak (blue) and $k_{B}T_{K}=95\protect\mu eV$ for the broad
Kondo peak (pink). b. (resp. c.) Comparison of scaled amplitude dependence
(resp. scaled frequency dependence) of $G_{ON}/G_{OFF}$ for broad (pink) and
narrow (blue) Kondo peaks to investigate universality. Squares correspond to
experimental data, while dots linked by solid lines correspond to the
Ansatz. }
\label{fig:universality}
\end{figure}
}

Finally, we investigate the crucial issue of the AC Kondo predicted
universality properties. Is $G_{ON}/G_{OFF}$ really governed by the reduced
parameters $eV_{AC}/k_{B}T_{K}$ and $hf/k_{B}T_{K}$ ? To address this
question, we compare the AC response of Kondo resonances with significantly
different Kondo temperatures. The peak fitted by the blue line in figure \ref%
{fig:universality}a is the "narrow" Kondo resonance, where the
frequency-amplitude characterization of previous section has been measured.
The peak fitted by the pink line is three times broader and corresponds to $%
T_{K}\sim 1.1K$. On the one hand, this factor 3 is advantageous, as scaling
should be tested unambiguously. On the other hand, it also means that the
investigated scaled frequency range is correspondingly reduced. Indeed our
microwave source delivers up to 20GHz frequencies, so we are limited to $%
hf/k_{B}T_{K}\lesssim 0.84$. Following the color code of \ref%
{fig:universality}a, figure \ref{fig:universality}b (resp. \ref%
{fig:universality}c) shows on the same graph $G_{ON}/G_{OFF}$ for the two
peaks as a function of scaled amplitude (resp. scaled frequency) at fixed
scaled frequency $hf/k_{B}T_{K}\approx 0.5$ (resp. scaled amplitude $e\tilde{%
V}_{AC}/k_{B}T_{K}\approx 2.3$). By definition, the pink and blue Ansatz
curves merge, up to small discrepancies due to noise in $G_{OFF}$ values. In
figure \ref{fig:universality}b, experimental data for both narrow and broad
Kondo peaks collapse well onto the Ansatz curve. This demonstrates the
universal dependence of $G_{ON}/G_{OFF}$ on ${V_{AC}}$ at low frequency. In
figure \ref{fig:universality}c, experimental data for the broad peak come
down to four points, as explained above. They fall reasonably well on the
narrow-peak data and confirm a flat adiabatic behavior at low scaled
frequency. This is consistent with a universal frequency dependence,
although universality in the pronounced frequency-dependant regime $%
hf/k_{B}T_{K}>1$ could not be investigated.

\section{CONCLUSION}

Our experiment realized on a CNT-based quantum dot brings conclusions to
several open issues of the AC Kondo problem. Using an independent in-situ
amplitude calibration, we provide the first quantitative measurement of the
amplitude-frequency dependence of the Kondo conductance with an AC bias. We
observe or not Kondo side peaks, depending on the excitation frequency used.
The Kondo resonance is found to be less affected at higher excitation
frequencies, in agreement with Kaminski et al. \cite{Kaminski2000}. We
describe our data quantitatively using an Ansatz, which bridges expressions
given in this paper for two limiting regimes of parameters. Measurement on
two Kondo resonances with very different $T_{K}$ allows to make tests of the
universality of the Kondo behavior. We find a good scaling of $%
G_{ON}/G_{OFF} $ with the scaled amplitude $eV_{AC}/k_{B}T_{K}$. In the more
limited parameter range we could explore, our data is also consistent with a
scaling of $G_{ON}/G_{OFF}$ with the scaled frequency $hf/k_{B}T_{K}$.

\section{Acknowledgements}
We thank Leonid Glazman for
fruitful discussions. We gratefully acknowledge J. Palomo,
M. Rosticher and A. Denis for technical support. The device has been made within
the consortium Salle Blanche Paris Centre. This work is supported by ERC Starting Grant CirQys.

\section{APPENDIX A: EXPERIMENTAL DETAILS}
\subsection{Sample fabrication and measurement} Carbon nanotubes (CNTs) were
grown on a separate quartz substrate containing mesas and transferred onto a
high-resistivity $Si$/$SiO_{2}$ substrate using a method described in ref. \cite{Viennot2014a}%
. After localizing CNTs with a Scanning Electron Microscope, the device
electrodes were patterned using electronic nanolithography. The two outer normal electrodes (70nm Pd) are connected to the
CNT with a room temperature resistance of $71k\Omega $. In contrast, the
central superconducting electrode (4nm Pd/100nm Al) is disconnected, and
acts as a top gate, on which we apply a microwave voltage to perform our
study of the AC Kondo. Transport measurements as a function of the two side
gates show parallel lines with a single slope, corresponding to a single dot
behaviour. Therefore a single side gate is operated with gate voltage $V_{g}$%
. In addition to the DC current, the differential conductance G is measured
by synchronous detection using a small AC bias voltage of 10$\mu V$ at 77.77
Hz. The device is embedded in a Nb coplanar waveguide resonator, which was
not used in this work as transmission measurements showed no coupling
between the two systems. \newline
\subsection{Fitting the Kondo resonance at equilibrium} We fit the Kondo
resonance conductance peak in absence of microwave excitation by an offset
Lorentzian curve (figures \ref{fig:sample}d and \ref{fig:universality}a) :
\begin{equation}
G=G_{BG}+\dfrac{G_{OFF}}{1+\dfrac{3}{8}\left( \dfrac{eV_{SD}-x_{c}}{%
k_{B}T_{K}}\right) ^{2}}  \label{equation:KondoAC:G_eq_fit}
\end{equation}%
The Lorentzian peak is centered on $x_{c}$ with an amplitude $G_{OFF}$ and a
width setting our practical definition of $T_{K}$. The parametrization is
chosen to recover the SU(2) coefficient of the (quadratic) Fermi liquid
description at low energy \cite{Glazman2003}. The presence of an offset
background conductance $G_{BG}$ is commonly observed in the Kondo regime
\cite{Jarillo-Herrero2005}. For instance, this could be a contribution of
cotunnelling, which would evolve over the scale $U\approx 1meV\gg k_{B}T_{K}$%
, which can be considered as flat over the Kondo resonance versus bias
voltage. \newline
\subsection{Microwave amplitude calibration} Our in-situ calibration method
consists in measuring the current rectification under microwave excitation
in the Coulomb regime. Considering our parameter range ($U\approx
1meV\gg hf$ and $\Gamma \approx 200\mu eV>2.5 hf$), we can reasonably assume that
the AC excitation simply modulates the DC current-voltage characteristic
I(V). This leads to the following expression of the measured current :
\begin{equation}
\underbrace{\left\langle I\left( V_{SD}+V_{AC}\cos \omega t\right)
\right\rangle }_{I_{ON}}=\underbrace{I\left( V_{SD}\right) }_{I_{OFF}}+%
\dfrac{1}{4}\frac{\mathrm{d^{2}}I}{\mathrm{d}V_{SD}^{2}}V_{AC}^{2}
\label{equation:KondoAC:rectification}
\end{equation}%
which is expressed to lowest order in the excitation voltage on the sample $%
V_{AC}$. This is valid if $V_{AC}$ explores the linear part of the
characteristic $G(V_{SD})$, leading to $V_{AC}\ll U\approx 1meV$. For each
frequency a proportionality coefficient c(f) links the sample voltage to the
source voltage : $V_{AC}=c(f)V_{AC,\text{source}}$. The c(f) values and the
detailed procedure to obtain them are given in the Appendix B
(figure 6). Note that equation \ref%
{equation:KondoAC:rectification} holds only if the microwave excitation
mainly acts as an AC bias voltage. This is the case of most excitation
frequencies, where the calibration coefficient c(f) is independent of the
gate voltage. Frequencies with a gate-dependent c(f) correspond to a mixed
AC bias-AC gate excitation regime, which is not the well controlled
situation we want to study. These frequencies are excluded from the
frequency set used for the AC Kondo characterization (for more details, see Appendix C and figure 7). \newline
\subsection{Scaled frequency-amplitude conductance map building procedure} The
experimental scaled frequency-amplitude conductance map shown in figure \ref%
{fig:FVmap}a (resp. in figure 9) is built from a database of
5189 sets (resp. 189) of (f, $V_{AC,\text{source}}$, G) gathering all
measurements of the microwave response of the "narrow" blue (resp. "broad"
pink) Kondo peak in figure \ref{fig:universality}a. The various measurement
types and data extraction procedures are detailed in the Appendices. The amplitude calibration coefficients c(f) are used to
calculate the excitation amplitude $V_{AC}$ from $V_{AC,\text{source}}$. $%
G_{BG}$, $G_{OFF}$ and $T_{K}$ are obtained from the equilibrium Kondo
resonance fit (see equation \ref{equation:KondoAC:G_eq_fit} above) before
each data acquisition (see table in figure 8 for parameters).
This allows to calculate the excitation scaled frequency $hf/k_{B}T_{K}$ and
scaled amplitude $eV_{AC}/k_{B}T_{K}$. The conductance under excitation $%
G_{ON}$ is obtained by subtracting the background conductance $G_{BG}$. We
assume $G_{BG}$ not being changed by the excitation, as $(hf,eV_{AC})\ll U$.
This assumption is validated by figure \ref{fig:ACphenomenology}c,d, where
conductance curves for all $V_{AC}$ values merge at high bias voltage. For
comparison between different data sets and/or between data and theory, it is relevant to normalize $G_{ON}$ by the
equilibrium conductance $G_{OFF}$ : $G_{ON}/G_{OFF}=\left( G-G_{BG}\right)
/G_{OFF}$. Finally, because data points are not uniformly spaced in $%
eV_{AC}/k_{B}T_{K}$-$hf/k_{B}T_{K}$ plane, we choose a regular grid and
obtain the map matrix by averaging data points in the same cell. Cells with
no data points remain blank. Here the cell size is 0.16 in $hf/k_{B}T_{K}$
and 0.1 in $eV_{AC}/k_{B}T_{K}$.
The Ansatz scaled frequency-amplitude conductance maps are built following the same procedure except that the values for $G_{ON}/G_{OFF}$ are calculated from the Ansatz formula \ref{equation:KondoAC:Ansatz} given in the main text, using experimental values for $eV_{AC}/k_{B}T_{K}$,$hf/k_{B}T_{K}$ and $G_{OFF}$. This explains why the Ansatz curves are not perfectly smooth. 

\section{Appendix B: In situ microwave amplitude calibration}

Our in-situ calibration method consists in measuring the current rectification under microwave excitation in the Coulomb regime. Considering our parameter range ($U\approx
1meV\gg hf$ and $\Gamma \approx 200\mu eV>2.5 hf$), we can reasonably assume that the AC excitation simply modulates the DC current-voltage characteristic I(V). This leads to the following expression of the measured current :
\begin{equation}
\label{equation:KondoAC:rectification}
\underbrace{\left\langle I\left(V_{SD}+V_{AC} \cos \omega t \right) \right\rangle}_{I_{ON}} =\underbrace{I\left(V_{SD} \right)}_{I_{OFF}} +\dfrac{1}{4}\frac{\mathrm{d^2}I }{\mathrm{d} V_{SD}^2}V_{AC}^{2}
\end{equation}
which is expressed to lowest order in the excitation voltage on the sample $V_{AC}$. This is valid if $V_{AC}$ explores the linear part of the characteristic $G(V_{SD})$, leading to $V_{AC}\ll U\approx 1meV$. For each frequency a proportionality coefficient c(f) links the sample voltage to the source voltage : $V_{AC}=c(f) V_{AC,\text{source}}$.

\begin{figure}\centering
\includegraphics[width=0.5\textwidth]{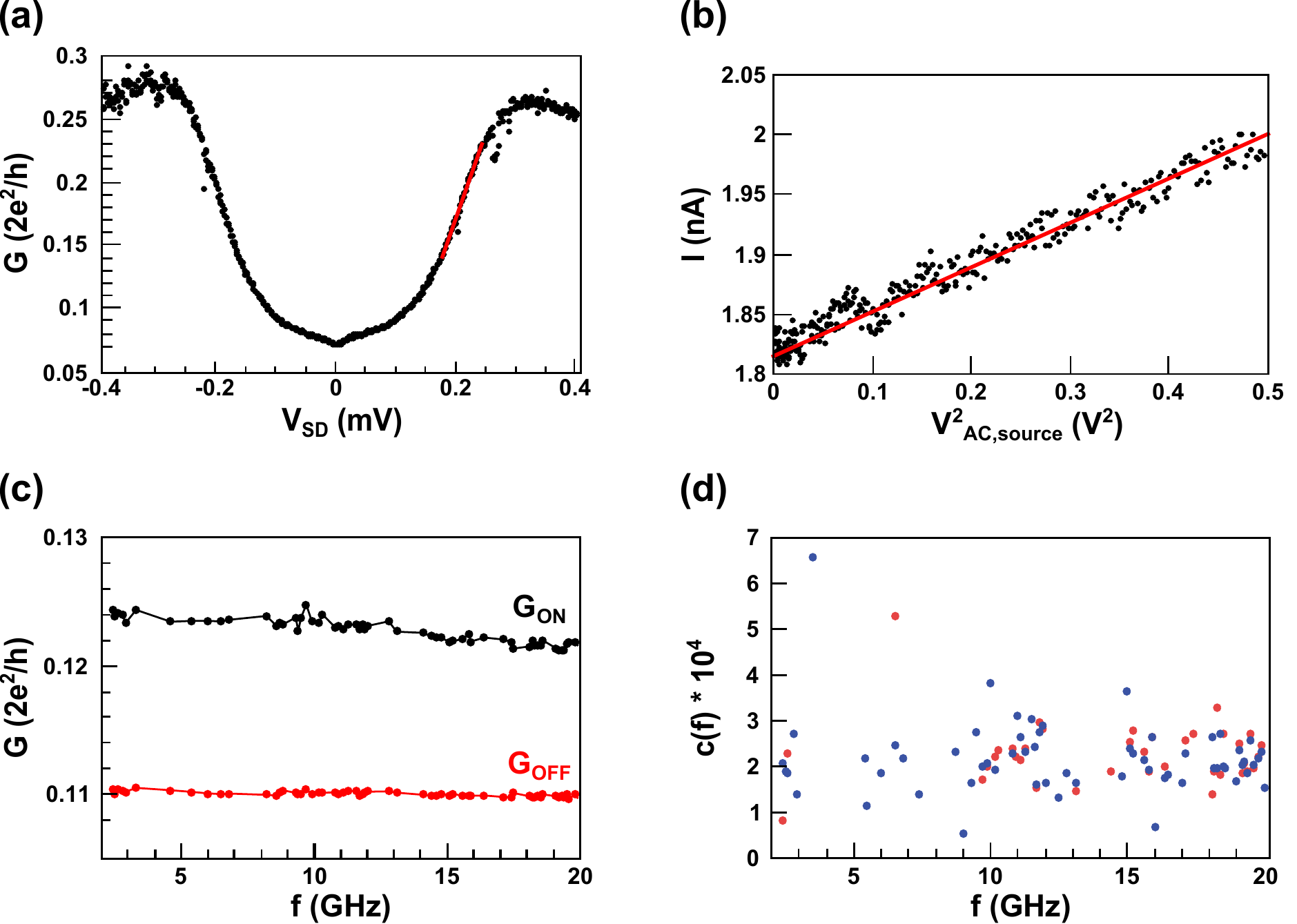}
\caption{\textbf{(a)} The local slope of the conductance G versus bias voltage gives $\frac{d^2 I}{d{V_{SD}^2}}$. Here around $V_{SD}=0.207$mV, $\frac{d^2 I}{d{V_{SD}^2}}=1.0 \, 10^{-1} A/V^{2}$. \textbf{(b)} Knowing this value, the slope p of the current I as a function of the square source amplitude $V_{AC,\text{source}}$ in the linear regime yields the amplitude calibration coefficient $c(f)=\sqrt{4p/\frac{d^2 I}{d{V_{SD}^2}}}$. Here for f=20GHz $p=0.372 nA/V^{2}$ hence $c(f=20GHz)=1.2\, 10^{-4}$.\textbf{(c)} shows an iterative relative amplitude calibration. For each frequency the conductance is successively measured with and without microwave excitation. Calibration coefficients $c_{n}(f)$ are used to set a $V_{AC}\approx 200 \mu V$ for each frequency. The new coefficient set $c_{n+1}(f)$ are deduced from equation \ref{equation:KondoAC:iterative_relative_cal} with G instead of I. \textbf{(d)} plots the calibration coefficients c(f) used to calculate the excitation amplitudes $V_{AC}$ applied to the narrow (blue dots) and broad (pink dots) Kondo peaks.}
\label{figure:KondoAC:calibration}
\end{figure}

Two calibration methods were combined to obtain the values of c(f). The first method provides a reliable absolute calibration. In absence of excitation ($V_{AC}=0$) $\frac{d^2 I}{d{V_{SD}^2}}$ is measured at $V_{SD,0}$ on the side of a Coulomb diamond. This is the local slope of the conductance as a function of bias voltage near $V_{SD,0}$, as depicted on figure \ref{figure:KondoAC:calibration}a. Figure \ref{figure:KondoAC:calibration}b shows the current at $V_{SD,0}$ as a function of $V_{AC}^{2}$. We focus on the low excitation regime, where the dependence is linear in agreement with \ref{equation:KondoAC:rectification}. Knowing $\frac{d^2 I}{d{V_{SD}^2}}$ at $V_{SD,0}$, the slope gives access to c(f). Applying this method for each frequency is time-consuming and the sample working point may change between first and last measured frequencies. Relative calibration is performed more rapidly and reliably using a second method. The source voltage is fixed and the frequency is swept. Current is successively measured with (ON current $I_{ON}$) and without (OFF current $I_{OFF}$) microwave excitation for each frequency point. For frequencies $f_1$ and $f_2$ with c(f) such, that the lowest-order expansion \ref{equation:KondoAC:rectification} is valid, calibration coefficients $c(f_{1})$ and $c(f_{2})$ are related by current rectification ratios :
\begin{equation}
\dfrac{c(f_{2})^{2}}{c(f_{1})^{2}}=\dfrac{I_{ON}(f_{2})-I_{OFF}}{I_{ON}(f_{1})-I_{OFF}}
\end{equation}
Relative calibration can be refined by iterating the procedure. We apply a source voltage $V_{AC,source}(f)=V_{AC,0}/c_{n}(f)$ calculated to apply a constant excitation voltage $V_{AC,0}\approx 200\mu V$. The new relative calibration is deduced from :
\begin{equation}
\label{equation:KondoAC:iterative_relative_cal}
\dfrac{c_{n+1}(f_{2})^{2}}{c_{n+1}(f_{1})^{2}}=\dfrac{c_{n}(f_{2})^{2}}{c_{n}(f_{1})^{2}} \dfrac{I_{ON}(f_{2})-I_{OFF}}{I_{ON}(f_{1})-I_{OFF}}
\end{equation}
This relation holds replacing the current I by the differential conductance G. Figure \ref{figure:KondoAC:calibration}c shows the conductance ON and OFF measurement, which defined the final relative calibration for some frequencies.

Figure \ref{figure:KondoAC:calibration}d shows the coefficients c(f) used to calibrate the amplitude $V_{AC}$ applied to the device for the study of the AC response of the narrow (blue dots) and broad (pink dots) Kondo peaks. Amplitude calibration for the narrow peak frequencies was done by combining absolute calibration and relative calibration methods described above. The resulting c(f) are on the same order of magnitude. This is because they have been preselected to be, otherwise relative powers may vary over tens of dB. We make this pre-selection to get a similar power range for all frequencies from the 43dB dynamic power range of our microwave source. For the broad Kondo peak, calibration was performed by iterating a relative calibration obtained on the narrow peak.
Calibration coefficients differ between broad and narrow peaks because a thermal cycling occurred between the two measurements runs. The thermal cycling did not change the calibration coefficients order of magnitude, but relative calibration was modified, possibly because wire bonds moved. Note that figure \ref{figure:KondoAC:calibration}d plots calibration coefficients only for frequencies kept after the frequency selection procedure, which we detail now.

\section{APPENDIX C: Frequency selection procedure}
\begin{figure}[!hpth]\centering
\includegraphics[width=0.45\textwidth]{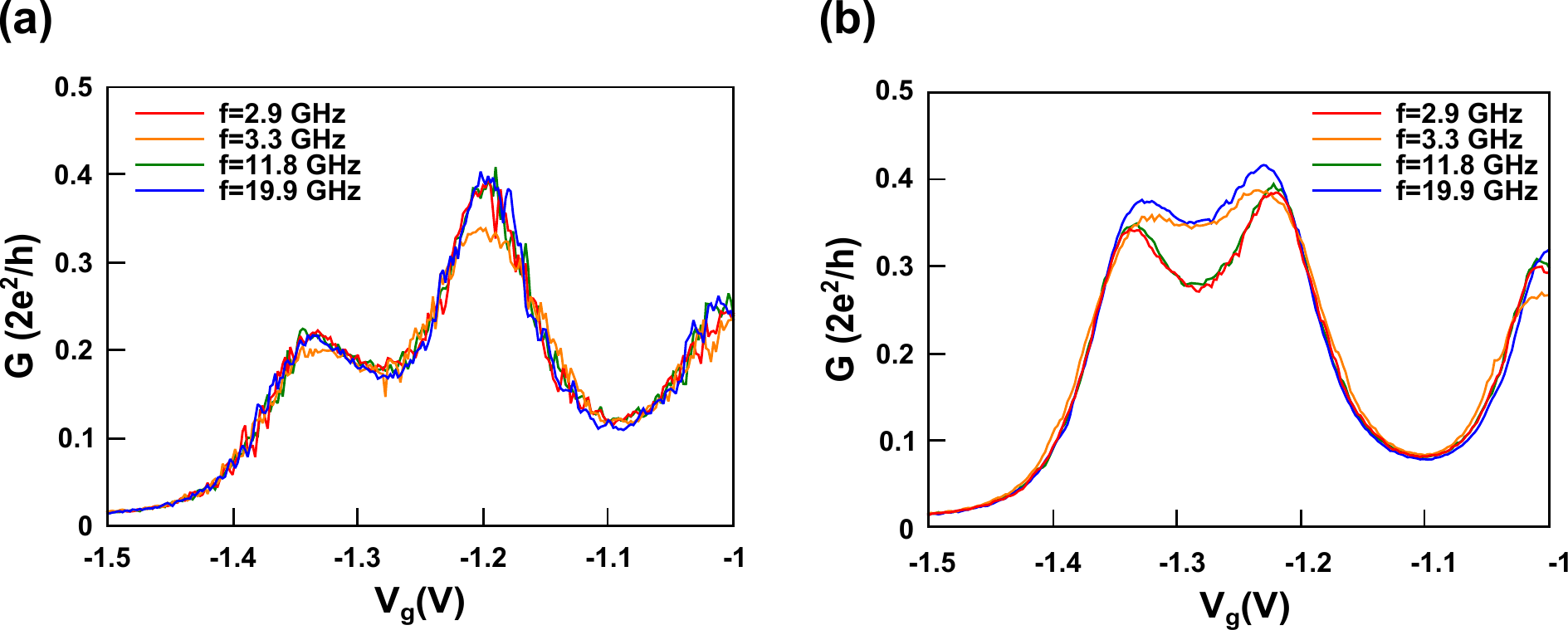}
\caption{\textbf{(a)} Conductance measurement at finite bias ($V_{SD}=203\mu V$) and under AC excitation $V_{AC}=100 \mu V$. Curves at f=2.9, 11.8 and 19.9 GHz superpose at all gate voltages, while the f=3.3 GHz does not. \textbf{(b)} Conductance measurement at zero bias under AC excitation ($V_{AC}=100 \mu V$) for the same frequencies.}
\label{figure:KondoACsup:cal_check}
\end{figure}

Once calibration is done, we study its gate dependence. We measure the conductance as a function of gate voltage successively with excitation ON ($V_{AC}=100 \mu V$) and OFF. This is done at finite bias $V_{SD}=203\mu V$ for all calibrated frequencies. As expected the curves coincide at $V_{g}=-1.095V$, where calibration was performed. We keep only frequencies, which curves superpose entirely. For example we observe on figure \ref{figure:KondoACsup:cal_check}a that frequencies 2.9, 11.8 and 19.9 GHz match well at all gate voltages, while 3.3 GHz does not. As a result 3.3 GHz is excluded from the study of AC Kondo. Frequencies remaining after this procedure correspond to an AC excitation mainly applied on the dot bias voltage. When current rectification is gate-dependent, the AC excitation very likely also applies to the dot energy level. Kaminski et al \cite{Kaminski2000} predicted an opposite frequency-dependence of the AC Kondo conductance with this kind of coupling, compared to coupling to bias. $G_{ON}/G_{OFF}$ is predicted to be decreasing with frequency instead of being increasing with frequency, as we measured. Figure \ref{figure:KondoACsup:cal_check}b shows that the Kondo conductance at 3.3 GHz is higher than at 2.9 GHz and 11.8 GHz, which is consistent with microwaves partly coupling to the gate. However we chose not to study this mixed coupling regime, as it is not easy to quantify couplings to bias and to gate separately. Note that the thermal cycling between the measurement runs affected the coupling mode at some frequencies, possibly because wire bonds moved. In the post-processing of the narrow peak data set, 25 frequencies were excluded starting from 86 frequencies. For the broad peak, 22 frequencies out of 59 were excluded.

\section{APPENDIX D: From raw data to scaled and normalized data}

\begin{figure}[!hpth]\centering
\includegraphics[width=1.00\linewidth]{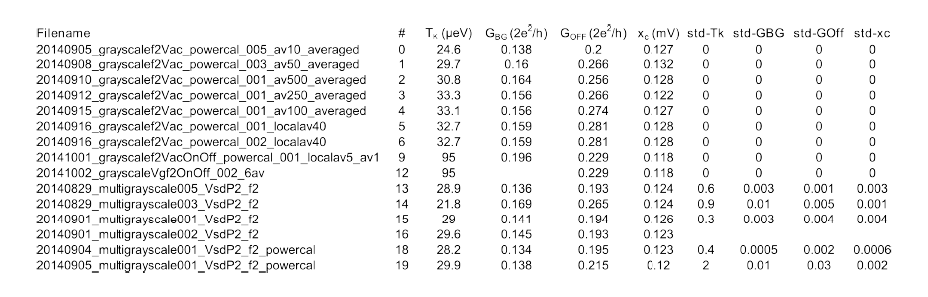}
\caption{Table of Kondo equilibrium parameters associated with each data file.}
\label{figure:KondoACsup:parameters_table}
\end{figure}

The AC response of a narrow Kondo peak ($T_{K}\approx 30 \mu eV$) was studied during a first measurement run. This corresponds to data files 0 to 6 ; 13 to 16 ; 18 and 19 in the table presented in figure \ref{figure:KondoACsup:parameters_table}. From these files 5189 sets of (f, $V_{AC,\text{source}}$, G) on the Kondo resonance have been extracted and used for the analysis.
After a thermal cycling the sample displayed a broad Kondo peak ($T_{K}\approx 95 \mu eV$). 189 data points were extracted from files 9 and 12 and used in the analysis.

Data types are explicitly indicated by the file name. They are three fold. Grayscale f2-Vac are measurements on the Kondo resonance peak, where the frequency is the sweeping loop, while the excitation voltage is the incremental loop. GrayscaleVgf2OnOff are gate sweeps for different frequencies at constant $V_{AC}$, the RF power being successively switched ON and OFF between sweeps. Multigrayscale-VsdP2-f2 are three loops measurements with the excitation frequency as external loop, the excitation voltage or power as intermediate loop and the bias voltage as internal loop.

We detail below how $G_{BG}$, $G_{OFF}$ and $T_{K}$ have been determined for each measurement type. The parameter values are tabulated in the table of figure \ref{figure:KondoACsup:parameters_table}. They were used to scale the excitation amplitudes and frequencies and normalise the conductance as described in the Appendix A.
\subsection{Grayscale f2-Vac}
The Kondo resonance at equilibrium is measured and fitted by an offset Lorentzian to obtain $G_{BG}$, $G_{OFF}$ and $T_{K}$ (see Appendix A). $V_{SD}$ and $V_{g}$ are fixed on the Kondo peak during the measurement as a function of excitation frequency and amplitude.
\subsection{GrayscaleVgf2OnOff}
For each gate sweep, we extract the locally minimal conductance value on the Kondo ridge.
This allows us to extract the Kondo conductance with and without microwave excitation for each frequency. As the Kondo conductance without excitation shifts slightly over time, we assume that $G_{BG}$ varies between sweeps, while $G_{OFF}$ is taken constant and is determined from fitting the Kondo resonance versus bias voltage without microwave excitation acquired before measuring GrayscaleVgf2OnOff.
\subsection{Multigrayscale-VsdP2-f2}
Here the equilibrium Kondo parameters are obtained for each frequency by fitting the conductance versus $V_{SD}$ measured at the lowest excitation amplitude, provided that $min(V_{AC})<15 \mu V$. This is justified by the observation that excitation amplitudes below this threshold do not affect the Kondo conductance compared to equilibrium. The  conductance values on the Kondo peak are extracted by filtering the points with $V_{SD}$ closest to the Lorentzian center fitting parameter $x_{c}$. $G_{ON}/G_{OFF}$, $hf/k_{B}T_{K}$ and $eV_{AC}/k_{B}T_{K}$ are calculated with the specific fitting parameters found for the corresponding frequency. Values tabulated in the table shown on figure \ref{figure:KondoACsup:parameters_table} are averaged quantities, with standard deviation indicated in other columns.

\section{APPENDIX E:Scaled amplitude-frequency conductance maps for the "broad" Kondo peak}
\begin{figure}[!hpth]\centering
\includegraphics[width=0.6\linewidth]{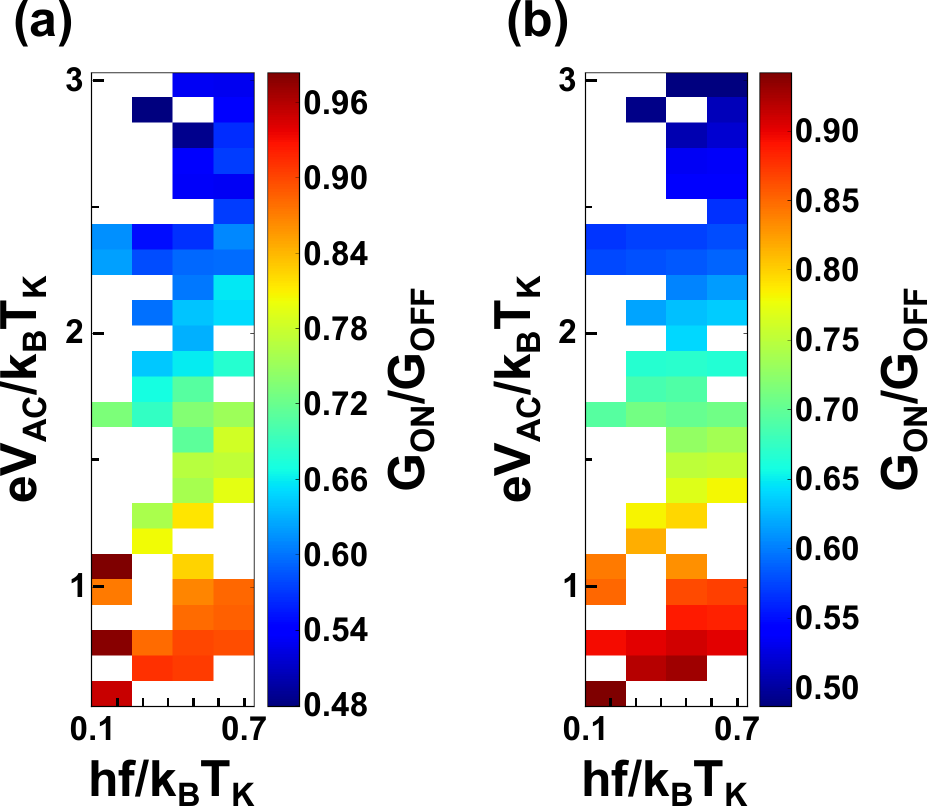}
\caption{Experimental \textbf{(a)} and Ansatz \textbf{(b)} scaled amplitude-frequency normalized conductance map for the broad peak.}
\label{figure:KondoACsup:FV_map_thick_peak}
\end{figure}

Figure \ref{figure:KondoACsup:FV_map_thick_peak} displays the experimental and Ansatz scaled amplitude-frequency normalized conductance maps for the broad Kondo peak. The higher number of blank cells compared to the "narrow" peak maps is due to the lower number of  data sets in the database from which the conductance matrix is built.

{\small 
}

\end{document}